\documentstyle[12pt,a4]{article}

\author{Afraimovich~E.~L., Altyntsev~A.~T.,
        Kosogorov~E.~A., Larina~N.~S., \and and Leonovich~L.~A. \\
        Institute of Solar-Terrestrial Physics SD RAS,\\
        p.~o.~box~4026, Irkutsk, 664033, Russia\\
        fax: +7 3952 462557; e-mail:~afra@iszf.irk.ru}

\title{Ionospheric effects of the solar flares of September 23,
1998 and July 29, 1999 as deduced from global GPS network data}

\date{}

\begin{document}
\sloppy
\maketitle

\begin{abstract}
This paper presents data from first GPS measurements of global
response of the ionosphere to solar flares of September 23, 1998
and July 29, 1999. The analysis used novel technology of a global
detection of ionospheric effects from solar flares (GLOBDET) as
developed by one of the authors (Afraimovich~E.~L.). The essence
of the method is that use is made of appropriate filtering and a
coherent processing of variations in total electron content (TEC)
in the ionosphere which is determined from GPS data,
simultaneously for the entire set of visible (over a given time
interval) GPS satellites at all stations used in the analysis. It
was found that fluctuations of TEC, obtained by removing the
linear trend of TEC with a time window of about 5 min, are
coherent for all stations and beams to the GPS satellites on the
dayside of the Earth. The time profile of TEC responses is
similar to the time behavior of hard X-ray emission variations
during flares in the energy range 25-35 keV if the relaxation
time of electron density disturbances in the ionosphere of order
50-100 s is introduced. No such effect on the nightside of the
Earth has been detected yet.
\end{abstract}

\section{Introduction}
\label{GLOB-sect-1}

The enhancement of X-ray and ultraviolet radiation intensity
that is observed during chromospheric flares on the Sun
immediately causes an increase in electron density in the
ionosphere. These density variations are different for
different altitudes and are collectively called Sudden
Ionospheric Disturbances (SID). SID observations provide a key
means for ground-based detection of solar flares along with
optical observations of flares and solar radio burst
observations. Much research is devoted to SID studies, among
them a number of thorough reviews and monographs \cite{Mit74}.

Unlike effects in the optical and radio ranges, ionospheric
effects of flares are of special interest as they constitute a
response of ionospheric plasma to an impulsive ionization.

Quantitative study of SID has two major implications at
present. SID data in the D-region that were obtained
predominantly by recording amplitude and phase characteristics
of signals from LF and VLF radio stations can be the source of
information about the X-ray part of the flare spectrum, as well
as providing a tool for investigating the principal chemical
processes in this region.

SID data for the $F$-region acquired by different radio probing
methods were used repeatedly to estimate time variations in the
X-ray and extreme ultraviolet (EUV) spectral regions and in
relative measurements of fluxes in different wavelength ranges
\cite{Don69}; \cite{Tho71}; \cite{Men74a}. The main body of SID
data for the Earth's upper atmosphere was obtained in earlier
detections of Sudden Frequency Deviations (SFD) of the
$F$-region-reflected radio signal in the HF range \cite{Dav69};
\cite{Don69}.

SFD are caused by an almost time-coincident increase in E-
and F-region electron densities at over 100 km altitudes
covering an area with the size comparable to or exceding that of
the region monitored by the system of HF radio paths. A
limitation of this method is the uncertainty in the spatial and
altitude localization of the UV flux effect, the inadequate
number of paths, and the need to use special-purpose equipment.

Another highly informative technique is the incoherent scatter
(IS) method, one of the most universal tools for ionosphere
research. In \cite{Tho71}, important information was obtained
about the height distribution of the decrease in local electron
density $N_e$ during the flares of May 21 and 23, 1967. A
significant increase of $N_e$ (by as much as 200 \%) was recorded
in the E-region, which decreased gradually in the F-region with
an up to 10-30 \% increase of the height and remained
distinguishable up to 300 km. $N_e$ starts to increase initially
in the E-region, while at higher altitudes it is observed to be
delayed, which is particularly conspicuous at the F-region
heights.

The Millstone Hill IS facility recorded the flare of August 7,
1972 \cite{Men74b}. The measurements were made in the range from
125 to 1200 km, i.e. to the altitudes exceeding greatly those of
all preceding observations. The increase of $N_e$ amounted to 100
\% at 125 km altitude and to 60 \% at 200 km.

Implementing the IS method requires extremely
sophisticated, expensive equipment. There are only a few IS
facilities world-wide, which are concentrated mainly in America
and Europe. These systems were designed for solving a broad
gamut of scientific problems and do not provide round-the-clock
observations of ionospheric effects from solar flares. An added
difficulty involves inadequate time resolution. Currently it is
common knowledge that the rise and fall time of the solar flare
emission in the range 10-1030 $\AA$, which has effect on the
ionospheric $…$- and $F$-regions, is often shorter than 5-10 min,
typical of the IS method's time resolution. Since the relaxation
time of electron density in the $E$- and $F1$-regions is also less
than 5-10 min, most incoherent scatter measurements lack
adequate time resolution for studying ionospheric effects of
flares.

The effect of solar flares on the ionospheric $F$-region is also
manifested as a Sudden Increase of Total Electron Content (SITEC)
which was measured previously using continuously operating VHF radio
beacons on geostationary satellites \cite{Mit74}; \cite{Men74a}.

In \cite{Men74a}, a pioneering attempt was made to realize global
observations of the outstanding flare of August 7, 1972 using 17
stations in North America, Europe, and Africa. The observations
covered a territory whose boundaries were separated by $70^{\circ}$
in latitude and by 10 hours in local time. For different stations, the
value of $dI$ (TEC increase) varied from $1.8 \,10^{16}$
¬${}^{-2}$ to $8.6 \,10^{16}$ ¬${}^{-2}$ , or $15-30\%$ of a total
electron content. These investigations revealed a latitudinal
dependence of the amount of TEC increase. The low latitudes showed
a larger increase of TEC compared with the high latitudes. Besides,
the authors point out no correlation between the value of TEC
increase and the solar zenith angle.

A limitation of the SITEC method is the integral character
of results which reflect the electron density variation in the
height range from 100 to 2000 km. If, however, it is taken into
consideration that only a few current methods enable flare
effects to be recorded in the ionospheric $F$-region, SITEC
observations should be recognized as one of the most convenient
tools for continuous observations of the $F$-region.

A serious limitation of methods based on analyzing VHF signals
from geostationary satellites is their small and ever increasing
(with the time) number and the nonuniform distribution in
longitude. Hence it is impossible to make measurements in some
geophysically interesting regions of the globe, especially in
high latitudes.

Consequently, none of the above-mentioned existing methods can
serve as an effective basis for the radio detection system to
provide a continuous, global SID monitoring with adequate
space-time resolution. Furthermore, the creation of these
facilities requires developing special-purpose equipment,
including powerful radio transmitters contaminating the radio
environment.

The advent and evolution of a Global Positioning System, GPS, and
also the creation on its basis of widely branched networks of GPS
stations (at least 600 sites at the end of 1999, the data from
which are placed on the INTERNET) opened up a new era in remote
ionospheric sensing. In the very near future this network will be
extended by integrating with the Russian navigation system,
GLONASS \cite{Klo97}. Furthermore, there exist also powerful
regional networks such as the Geographical Survey Institute
network in Japan \cite{Sai98} consisting of up to 1000 receivers.
High-precision measurements of the group and phase delay along
the line-of-sight (LOS) between the receiver on the ground and
transmitters on the GPS system satellites covering the reception
zone are made using two-frequency multichannel receivers of the
GPS system at almost any point of the globe and at any time
simultaneously at two coherently coupled frequencies
$f_1=1575{.}42$ MHz and $f_2=1227{.}60$ MHz.

The sensitivity of phase measurements in the GPS system is
sufficient for detecting irregularities with an amplitude of up
to $10^{-3}$--$10^{-4}$ of the diurnal TEC variation. This makes
it possible to formulate the problem of detecting ionospheric
disturbances from different sources of artificial and natural
origins. Recently some authors embarked actively on the
development of detection tools for the ionospheric response of
powerful earthquakes \cite{Cal95}, rocket launches \cite{Cal96},
and industrial surface explosions \cite{Fit97}; \cite{Cal98}.
Subsequetly, the GPS data began to be used in the context of the
spaced-receiver method using three GPS stations to determine the
parameters of the full wave vector of traveling ionospheric
disturbances under quiet and disturbed geomagnetic conditions
\cite{Afr98}; \cite{Afr99}.

The objective of this paper is to develop a method of global
detection of the ionospheric effect from solar flares (GLOBDET)
using the international GPS network. This method would improve
substantially the sensitivity and space-time resolution of
analysis when compared with the above-mentioned radio probing
methods. General information about the flares being analyzed here
and a description of the experimental geometry are given in
Section ~\ref{GLOB-sect-2}. The processing technique for the data
from the GPS network and results of an analysis of the
ionospheric effect from the solar flares of September 23, 1998
and July 29, 1999 are outlined in Section ~\ref{GLOB-sect-3}.
Section ~\ref{GLOB-sect-4} discusses the results obtained. A
modeling of the physical processes involving flare effects on the
ionosphere using GPS data and, moreover, the development of
methods for solving their inverse problem of reconstructing
emission characteristics using GPS data will be subject of future
research.

\section{ Experimental geometry, and general data on the solar
flares of September 23, 1998 and July 29, 1999}
\label{GLOB-sect-2}

For studying the ionospheric response to the ionizing emission
from solar flares, we chose relatively powerful (according to an
X-ray classification) flares whose time profile was
characterized by intense short-duration impulses of hard X-ray
emission. The two flares selected appear in Table 1.
The table also provides information about the
characteristics of the CGRO- and YOHKOH-borne X-ray detectors,
the data from which are compared with the ionospheric response
to these flares in this paper.

Fig. 1 presents the geometry of a global GPS array used in this paper
to analyze the effects of flare 23 September 1998 (102 stations -- a.)
and 29 July 1999 (105 station--b.). Heavy dots correspond to the
location of the GPS stations. The coordinates of the stations are not
given here for reasons of space. The upper scales indicate the local
time, LT, corresponding to 07:00 UT, a maximum increase in X-ray
emission intensity of the flare 23 September 1998 and 19:30 UT for
29 July 1999 (see Section ~\ref{GLOB-sect-3})).

As is evident from Fig.1, the set of stations which we chose out
of the global GPS network available to us, covers rather densely
North America and Europe, but provides much worse coverage of the
Asian part of the territory used in the analysis. The number of
GPS stations in the Pacific and Atlantic regions is even fewer.
However, coverage of the territory with partial beams to the
satellite for the limitation on elevations $\theta>10^\circ$~,
which we have selected, is substantially wider. Dots in Fig. 1c
mark the coordinates of subinospheric points for the height of
the $F2$--layer maximum $h_{max}=300$ km for all visible
satellites at 29 July 1999, 19:30 UT for each GPS station. A
total number of beams (and subionospheric points) used in this
paper to analyze the July 29, 1999 flare is 622.

Such coverage of the terrestrial surface makes it possible to
solve the problem of detecting time-coincident events with
spatial resolution (coherent accumulation) two orders of
magnitude higher, as a minimum, than could be achieved in SFD
detection on oblique HF paths. For simultaneous events in the
western hemisphere, the correspoding today's number of stations
and beams can be as many as 400 and 2000--3000, respectively.

Figs. 2b and 3b show the time dependencies of flare
emission. Time profiles of soft X-ray emission were acquired by
the GOES-10 satellite, the  data from which
are available on the SPIDER network (marked by the symbol o).
Dashes in Figs. 2b and 3b represent values for the low-energy
1-8$\AA$ channel. The time profile of the signal in the 0.5-4$\AA$
channel was alike. A comparative analysis of these series is
made in Section 6.

The study was based on events, for which flare emission data
with a time resolution of about one second were available. The
September 23, 1998 flare was recorded by X-ray telescope HXT on
the YOHKOH satellite (Fig.~3b). Operating in the flare mode, the
HXT telescope provides observations in four energy channels
(14-23 keV, 23-33 keV, 33-53 keV, and 53-93 keV) with a
resolution of 0.5 s. The second event was observed by the BATSE
spectrometer on the CGRO satellite which is capable of recording
solar X-ray emission with different temporal and spectral
resolutions. This study utilized DISCLA data written in four
channels (25-50 keV, 50-100 keV, 100-300 keV, and over 300 keV)
at 1.024-second intervals.

The events are both characterized by a low level of
geomagnetic disturbance (from -10 to -20 nT), which simplified
greatly the SID detection problem.

\section{Processing of the data from the GPS network, and
results derived from analyzing the ionospheric effect from the
solar flares of September 23, 1998 and July 29, 1999}
\label{GLOB-sect-3}

Following is a brief outline of the global monitoring (detection)
technique for solar flares (GLOBDET) as developed by one of the
authors (Afraimovich~E.~L.) on the basis of processing the data
from a worldwide network of two-frequency multichannel receivers
of the GPS-GLONASS navigation systems.

A physical groundwork for the method is formed by the effect of
fast change in electron density in the Earth's ionosphere at the
time of a flare simultaneously on the entire sunlit surface.
Essentially, the method implies using appropriate filtering and a
coherent processing of TEC variations in the ionosphere
simultaneously for the entire set of "visible" (during a given
time interval) GPS satellites (as many as 5-10 satellites) at all
global GPS network stations used in the analysis. In detecting
solar flares, the ionospheric response is virtually simultaneous
for all stations on the dayside of the globe within the time
resolution range of the GPS receivers (from 30 s to 0.1 s).
Therefore, a coherent processing of TEC variations implies in
this case a simple addition of single TEC variations.

The detection sensitivity is determined by the ability to detect
typical signals of the ionospheric response to a solar flare
(leading edge duration, period, form, length) at the level of TEC
background fluctuations. Ionospheric irregularities are
characterized by a power spectrum, so that background
fluctuations will always be distinguished in the frequency range
of interest. However, background fluctuations are not correlated
in the case of beams to the satellite spaced by an amount
exceeding the typical irregularity size.

With a typical length of X-ray bursts and EUV emission of solar
flares of about 5-10 min, the corresponding ionization
irregularity size does normally not exceed 30-50 km; hence the
condition of a statistical independence of TEC fluctuations at
spaced beams is almost always satisfied. Therefore, coherent
summation of responses to a flare on a set of beams spaced
thoughout the dayside of the globe permits the solar flare effect
to be detected even when the response amplitude on partial beams
is markedly smaller than the noise level (background
fluctuations). The proposed procedure of coherent accumulation is
essentially equivalent to the operation of coincidence schemes
which are extensively used in X-ray and gamma-ray telescopes.

If the SID response and background fluctuations, respectively,
are considered to be the signal and noise, then as a consequence
of a statistical independence of background fluctuations the
signal/noise ratio when detecting the flare effect is increased
through a coherent processing by at least a factor of $\sqrt{N}$,
where $N$ is the number of LOS.

It should be noted that because of the relatively low satellite
orbit inclinations, the GPS network (and to a lesser degree
GLONASS) provides poor coverage of the Earth's surface near the
poles. However, TEC measurements in the polar regions are
ineffective with respect to the detection of the ionospheric
response to a solar flare because the amplitude of background
fluctuations in this case is much higher when compared with the
mid-latitude ionosphere. This is partiocularly true of
geomagnetic disturbance periods. For the same reason, equatorial
stations should also be excluded from a coherent processing.

The GPS technology provides the means of estimating TEC
variations on the basis of phase measurements of TEC $I$ in each
of the spaced two-frequency GPS receivers using the formula
\cite{Hof92}; \cite{Cal96}:

\begin{equation}
\label{GLOB-eq-01}
I=\frac{1}{40{.}308}\frac{f^2_1f^2_2}{f^2_1-f^2_2}
                           [(L_1\lambda_1-L_2\lambda_2)+const+nL]
\end{equation}

where $L_1\lambda_1$ and $L_2\lambda_2$~ are phase path
increments of the radio signal, caused by the phase delay in the
ionosphere (m); $L_1$, $L_2$~ is the number of full phase
rotations, and $\lambda_1$, and $\lambda_2$, are the wavelengths
(m) for the frequencies $f_1$ and $f_2$, respectively; $const$~
is some unknown initial phase path (m); and $nL$~ is the error in
determination of the phase path (m).

Phase measurements in the GPS system are made with a high degree
of accuracy where the error in TEC determination for 30-second
averaging intervals does not exceed $10^{14}$ ¬${}^{-2}$,
although the initial value of TEC does remain unknown
\cite{Hof92}. This permits ionization irregularitues and wave
processes in the ionosphere to be detected over a wide range of
amplitudes (as large as $10^{-4}$ of the diurnal variation of
TEC) and periods (from several days to 5 min). The TEC unit,
$TECU$, which is equal to $10^{16}$ ¬${}^{-2}$ and is commonly
accepted in the literature, will be used throughout the text.

The solar flare of July 29, 1999 was used to illustrate the
performance of the proposed method.
Primary data include series of "oblique" values of
TEC $I(t)$, as well as the corresponding series of elevations
$\theta(t)$ and azimuths $\alpha(t)$ along LOS to the
satellite calculated using our developed CONVTEC program which
converts the GPS system standard RINEX-files on the INTERNET
\cite{Gur93}. The determination of SID characteristics involves
selecting continuous series of $I(t)$ measurements of at least a
one-hour interval in length, which includes the time of the
flare. Series of elevations $\theta(t)$ and azimuths
$\alpha(t)$ of the beam to the satellite are used to determine
the coordinates of subionospheric points. In the case under
consideration, all results were obtained for elevations
$\theta(t)$ larger than $10^\circ$.

Fig.~4a presents typical time dependencies of an "oblique" TEC
$I(t)$ for the PRN03 satellite at the CME1 station on July 29,
1999 (thick line) and for PRN21 at the CEDA station (thin line).
It is apparent from Fig.~4a that in the presence of slow TEC
variations, the SID-induced short-lasting sudden increase in TEC
is clearly distinguished in the form of a "step" as large as 0.4
$TECU$.

For the same series, similar lines in panel b. show variations of
the time derivative of TEC $dI(t)/dt$ with the linear trend
removed and with a smoothing with the 5-min time window. The TEC
time derivative is invoked because it reflects electron density
variations which are proportional to the X-ray or EUV flux
\cite{Mit74}.

The $dI(t)/dt$ variations for different beams are well correlated
over the time interval from 19:30 to 19:39 UT. This is
distinguished in a more instructive way if series of the time
derivative $dI(t)/dt$ for all visible satellites from 105 GPS
stations are plotted on the same time scale (panel d).
Time-coincident (for all beams) $dI(t)/dt$ variations are clearly
seen; for the remaining time spans of the time interval
19:00--20:00, UT variations of $dI(t)/dt$ for different sites and
satellites are not correlated and occupy the entire amplitude-time
range.

The coherent summation of $dI(t)/dt_i$ realizations was made by the
formula

\begin{equation}
\label{GLOB-eq-02}
\Sigma dI(t)/dt = \sum^N_{i=1} dI(t)/dt_i sin(\theta_i)
\end{equation}

where $\theta_i$ is LOS elevation, $i$ - number of LOS; $i=$ 1,
2, ... $N$.

Multiplication by $sin(\theta_i)$ was used to convert "oblique" TEC
variations to an "equivalent" vertical value in order to
normalize the response amplitude.

The normalized to $N$ result of a coherent summation
~(\ref{GLOB-eq-02}) for all beams and GPS stations located mainly
on the dayside is presented in panel c. A comparison of the
resulting coherent sum ~(\ref{GLOB-eq-02}) with the time
dependence $dI(t)/dt$ for individual beams presented in panels b.
and d. confirms the effect of a substantial increase of the
signal/noise ratio caused by a coherent processing.

It is interesting to compare, for the same time interval, the
data on individual beams and results from a coherent summation
for the dayside and nightside. Fig.~4e presents typical time
dependencies of an "oblique" TEC $I(t)$ for the PRN27 satellite
at the IRKT station (thick line) and for PRN27 at the BOGO
station (thin line). Using the $I(t)$ data it is impossible to
identify any SID-induced short-lasting sudden increase in TEC.
This is also true for the time derivatives $dI(t)/dt$ plotted in
panels f. and h. As a result, the r.m.s. of the coherent sum
~(\ref{GLOB-eq-02}) in panel g. for the nightside is of the same
order of magnitude as that of background fluctuations outside the
SID response interval on the dayside, which is an order of
magnitude (as a minimum) smaller than the SID response amplitude
(Fig.~4c).

Consider the data processing procedure for the solar flare of
September 23, 1998. Fig. 5 presents the time dependencies of TEC
$I(t)$ on the dayside -- a. and $dI(t)$ variations with the
linear trend removed and smoothing with the 5-min time window for
stations IRKT (PRN01) -- b; $dI(t)$ variations for all visible
satellites from 102 GPS stations plotted on the same time scale --
d. One can clearly see simultaneous (for all GPS beams) $dI(t)$
variations; for the remaining time spans of the interval
6:30-08:00 UT selected, $dI(t)$ variations for different sites
and satellites are not correlated and occupy the entire
amplitude-time range.

It should be noted that in this case the ratio of the amplitude of the
ionospheric response of the flare to the phase fluctuation amplitude is
substantially worse than that for the flare of July 29, 1999. A
preliminary coherent accumulation of $dI(t)_i$-series with a
subsequent differentiation, rather than differentiation of single
realizations of $dI(t)_i$ with subsequent addition of the time
derivatives seems more reasonable. The coherent summation of
$dI(t)_i$ realizations was made for this event by the formula

\begin{equation}
\label{GLOB-eq-03}
\Sigma dI(t) = \sum^N_{i=1} dI(t)_i sin(\theta_i)
\end{equation}

The (normalized to $N$) result of the coherent summation of series
~(\ref{GLOB-eq-03}) for all beams and GPS stations located
predominantly on the dayside is presented in panel c. Again, a
comparison of the resulting coherent sum ~(\ref{GLOB-eq-03}) with
the time dependence of $dI(t)$ for the individual beams shown in
panels b. and d. confirms the enhancement effect of the signal/noise
ratio beause of a coherent processing.

A similar result is also obtained by comparing (for the same time
interval) the data from individual LOS and coherent summation results
for the dayside and nightside. Fig.~5c presents typical time
dependencies of an "oblique" TEC $I(t)$ for the PRN14 satellite at
station MAS1. The SID-induced short-lasting sydden increasze of
TEC was not possible to identify from the $I(t)$ data. This applies
also for the $dI(t)$-variations with the trend removed, which are
plotted in panels~f. and h. As a result, the standard deviation (SD) of
the coherent sum ~(\ref{GLOB-eq-03}) in panel g. for the nightside
is found to be of the same order o magnitude as the backgrund
fluctuation SD outside the SID response range on the dayside, which
is an order of magnitude, as a minimum, less than the SID response
amplitude (Fig.~5c).

\section{Discussion}
\label{GLOB-sect-4}
A comparative analysis is made of the TEC data and X-ray
emission time series acquired by satelites. In carrying out a
comparative analysis of the TEC  and X-ray emission to
flares, it is necessary to eliminate the TEC trend which is not
associated with flare emission. In this case the above procedure
of determining the trend that is removed by a smoothing over a
5-min interval, which is significantly shorter than the flare
emission duration, is incorrect. In Figs. 2 and 3, the trend to
be removed was therefore defined as a polynomial of degree 3,
approximating the time dependence of TEC on the intervals 19:00
and 20:00 UT, and 6:00-8:00 UT, respectively. The approximation
procedure neglected the TEC values during the flares
(19:30-19:48 UT, and 6:40-7:24 UT).

First we consider a simpler flare of July 29, 1999, with the
X-ray emission time profile like a single impulse (dashed curve
$F_{HXT}$ in Fig.~2b). The flare is clearly seen on the time
profile of TEC $I(t)$ (Fig.~2a). As is apparent from Fig.~2b, the
response $I_{ex}(t)$ represents an impulse with a fast growth and
a relatively slow decline. The time variation of soft X-ray
emission, obtained from the GOES data at insufficiently large
time intervals (marked by the symbol o in Fig.~2b) does not
contradict the behavior of the curve $I_{ex}(t)$.

The rise front of hard X-ray emission $F_{HXT}$ is steeper when
compared with the response of TEC $I_{ex}(t)$ , and the time of a
maximum is 1.5 min ahead of the TEC fluctuation maximum. These
difference are natural if account is taken of the finite relaxation time
of electron density disturbances caused by flare emission. This factor
can be taken into account by convoluting the source function
$F_{HXT}$ with the relaxation function:

\begin{equation}
\label{GLOB-eq-04}
F_{conv}(t)=
\int\limits_{0}^{t} F_{\tiny\sf HXT}(t')
\exp\left[-\left(\frac{t-t'}{\tau}\right)\right]dt'
\end{equation}

As is evident from Fig.~2c, the growth profile of TEC $I_{ex}(t)$
is similar to that of the convolution $F_{conv}$ when $\tau=$ 65
s. The convolution was accomplished with the X-ray signal of the
25-50 keV (0.25-0.5 A) energy channel. At the decay phase, the
curves $F_{conv}$ and $I_{ex}(t)$ are moving apart, which can be
associated with the contribution to the ionospheric ionization
from softer emission whose time profile is similar to the GOES
flux profile.

Short-duration disturbances on TEC dependencies, which are
associated with the ionization by flare emission, are more
pronounced on the time derivatives $dI(t)/dt$. Fig.~2d compares
the result of summation ~(\ref{GLOB-eq-02}) of the series of the
derivatives $\Sigma dI(t)/dt$ for all visible satellites from GPS
stations located on the dayside, with the derivative
$d/dt(F_{conv})$ presented in Fig. 2c. It can be seen that the
growth stage of the coherent sum $\Sigma dI(t)/dt$ for the entire
sunlit side of the Earth is described adequately by the function
$d/dt(F_{conv})$. Note that the accuracy of estimating the
duration $\tau$ determined from the coincidence of peaks of the
time derivatives is determined by time resolution of TEC
measurements (30~s in the case under consideration).

The flare of September 23, 1998 was of a longer duration, and,
despite a somewhat higher intensity, its response was relatively
small on TEC time dependencies for separate paths (Fig.~3a).
Nevertheless, by subtracting the polynomial of degree 3, it was
possible to identify the TEC response $I_{ex}(t)$ (Fig.~3b). In
this event, the time dependence of soft X-ray emission is much
different from the temporal behavior of the response $I_{ex}(t)$
which grows faster than does the GOES signal (marked by the
symbol o), attains a maximum 9 min earlier, and decreaes much
more rapidly.

The signal of hard X-ray emission $F_{HXT}$ (dashed line) shows a
number of peaks leading the TEC fluctuations $I_{ex}(t)$. The
convolution with the hard X-ray emission signal $F_{conv}$ agrees
satisfactorily with the response $I_{ex}(t)$, with $\tau=$ 100 s. The
curves in Fig. 2c are most similar for the M1 (22-35 keV) channel of
the HXT/YOHKOH X-ray telescope. The estimated $\tau=$ 100 s is
confirmed by comparing the combined time derivative of TEC
$\Sigma dI(t)/dt$ , with the derivative $d/dt(F_{conv})$.

A comparison of the time profiles of the TEC response and hard
X-ray emission shows that corresponding electron density
disturbances are recorded with confidence by a global GPS network.
For hard X-ray emission, the highest correlation is attained for
photon energies of about 30 keV, with relation times $\tau$ in the
range 65-100 s. These estimates are in reasonably good agreement
with results obtained previously when analyzing the SID effect
\cite{Don69}, \cite{Mit74}.

Unfortunately, the lack of data on UV emission of the flares
under investigation, the most probable ionizing factor at
ionospheric heights above 100 km, gives no way of making absolute
estimates of the TEC increment and comparing them with measured
values. It is pointed out in \cite{Mit74} that although UV
emission is essentially responsible for SID in the F-region, TEC
variations are also correlated quite well with X-ray flares. This
is also confirmed by simultaneous measurements of X-ray and EUV
flare emission characteristics by the Solar Maximum Mission
satellite \cite{Van88}.

\section{Conclusions}
\label{GLOB-sect-5}

In this paper we have analyzed the ionospheric response to
powerful solar flares of September 23, 1998 and July 29, 1999.
The analysis is based on implementing our new technology of a
global detector of ionospheric effects from solar flares using
the data from the international network of two-frequency
multichannel receivers of the navigation GPS system (GLOBDET)
which improves substantially the sensitivity and space-time
resolution of observations over existing radio probing methods.

It was found that fluctuations of TEC and its time derivative
obtained by removing the linear trend of TEC with a time window
of about 5 min are coherent for all stations and beams to GPS
satellites on the dayside of the Earth, regardless of the
station's location, local time and elevation of the beam to the
satellite. The time profile of TEC responses is similar to the
time behavior of hard X-ray emission variations during flares in
the energy range 25-35 keV if we introduce the relation time of
an electron density disturbance in the ionosphere of a duration
of 50-100 s. No such effect was detected on the nightside of the
Earth.

The GLOBDET technology, suggested in this paper, can be used to
detect small solar flares; the body of data processed is the only
limitation in this case. The high sensitivity of GLOBDET permits
us to propose the problem of detecting, in the flare X-ray and
EUV ranges, emissions of nonsolar origins which are the result of
supernova explosions.

For powerful solar flares like the one examined in this report,
it is not necessary to invoke a coherent summation, and the SID
response can be investigated for each beam. This opens the way to
a detailed study of the SID dependence on a great variety of
parameters (latitude, longitude, solar zenith angle, spectral
characteristics of the emission flux, etc.). With current
increasing solar activity, such studies become highly
challenging. In adidtion to solving traditional problems of
estimating parameters of ionization processes in the ionosphere
and problems of reconstructing emission parameters \cite{Mit74},
the data obtained through the use of GLOBDET can be used to
estimate the spatial inhomogeneity of emission fluxes at scales
of the Earth's radius.

\section*{Acknowledgments}
\label{GLOB-sect-6}

Authors are grateful to V.~V.Grechnev,~E.~A. Kosogorov,~O.~S.
Lesuta and ~K. ~S. Palamartchouk for preparing the input data.
Thanks are also due V.~G.~Mikhalkovsky for his assistance in
preparing the English version of the \TeX-manuscript. This work
was done with support from the Russian Foundation for Basic
Research (grants 97-02-96060 and 99-05-64753), GNTP 'Astronomy'
as well as RF Minvuz Grant 1999; supervisor B.O. Vugmeister.

{}
\end{document}